\documentclass[twocolumn,aps,prc,showpacs,superscriptaddress,preprintnumbers,floatfix,nofootinbib,longbibliography]{revtex4-1}
\usepackage{multirow}
\usepackage{hhline}
\usepackage{bbm}
\usepackage{epsfig}
\usepackage{graphics}  % needed for figures
\usepackage{graphicx}  % needed for figur\es
\usepackage{dcolumn}   % needed for some tables
\usepackage{bm}        % for math
\usepackage{amssymb}   % for math
\usepackage{amsmath}   % for math
\usepackage{amsfonts}
\usepackage[percent]{overpic}
\usepackage{natbib}
\usepackage{hyperref}% add hypertext capabilities
\usepackage[usenames]{color}
\usepackage{CJK}

\usepackage{tikz,xcolor,hyperref}

\definecolor{lime}{HTML}{A6CE39}
\DeclareRobustCommand{\orcidicon}{
	\begin{tikzpicture}
	\draw[lime, fill=lime] (0,0) 
	circle [radius=0.16] 
	node[white] {{\fontfamily{qag}\selectfont \tiny ID}};
	\draw[white, fill=white] (-0.0625,0.095) 
	circle [radius=0.007];
	\end{tikzpicture}
	\hspace{-2mm}
}
\foreach \x in {A, ..., Z}{%
	\expandafter\xdef\csname orcid\x\endcsname{\noexpand\href{https://orcid.org/\csname orcidauthor\x\endcsname}{\noexpand\orcidicon}}
}
\foreach \x in {A, ..., Z}{%
	\expandafter\xdef\csname orcid\x\endcsname{\noexpand\href{https://orcid.org/\csname orcidauthor\x\endcsname}{\noexpand\orcidicon}}
}

\begin{document}

\title{Validation and extrapolation of atomic mass with physics-informed fully connected neural network}

\newcommand{\moe}{Key Laboratory of Nuclear Physics and Ion-beam Application (MOE), and Institute of Modern Physics, Fudan
University, Shanghai 200433, China}
\newcommand{\fudan}{Shanghai Research Center for Theoretical Nuclear Physics, NSFC and Fudan University, Shanghai 200438, China}
\newcommand{\jilin}{College of Physics, Jilin University, Changchun 130012, China}
\newcommand{\fudanP}{Physics Department and Center for Particle Physics and Field Theory, Fudan University, Shanghai 200438, China}
\newcommand{\sbu}{Department of Chemistry, Stony Brook University, Stony Brook, NY 11794, USA}
\newcommand{\bnl}{Physics Department, Brookhaven National Laboratory, Upton, NY 11976, USA}
\author{Yiming Huang\orcidA{}}\affiliation{\moe}\affiliation{\fudan}\affiliation{\jilin}
\author{Jinhui Chen\orcidB{}}\email{chenjinhui@fudan.edu.cn}
\affiliation{\moe}\affiliation{\fudan}
\author{Jiangyong Jia\orcidC{}}\email{jiangyong.jia@stonybrook.edu}
\affiliation{\sbu}\affiliation{\bnl}
\author{Lu-Meng Liu\orcidD{}}\email{liulumeng@fudan.edu.cn}\affiliation{\fudanP}
\author{Yu-Gang Ma\orcidE{}}\email{mayugang@fudan.edu.cn}\affiliation{\moe}\affiliation{\fudan}
\author{Chunjian Zhang\orcidF{}}\email{chunjianzhang@fudan.edu.cn}\affiliation{\moe}\affiliation{\fudan}\affiliation{\sbu}

\begin{abstract}
Machine learning offers a powerful framework for validating and predicting atomic mass. We compare three improved neural network methods for representation and extrapolation for atomic mass prediction. The powerful method, adopting a macroscopic-microscopic approach and treating complex nuclear effects as output labels, achieves superior accuracy in AME2020, yielding a much lower root-mean-square deviation of 0.122 MeV in the test set, significantly lower than alternative methods. It also exhibits a better extrapolation performance when predicting AME2020 from AME2016, with a root-mean-square deviation of 0.191 MeV. We further conduct sensitivity analyses against the model inputs to verify interpretable alignment beyond statistical metrics. Incorporating theoretical predictions of magic numbers and masses, our fully connected neural networks reproduce key nuclear phenomena including nucleon pairing correlation and magic number effects. The extrapolation capability of the framework is discussed and the accuracy of predicting new mass measurements for isotope chains has also been tested.
\end{abstract}
%\doi{}
\maketitle
\section{Introduction}
\label{sec:intro}
The atomic nucleus, composed of nucleons (protons and neutrons), represents a complex quantum many-body system~\cite{Lunney:2003zz,Martin:2015xql}. Its mass is a fundamental property critical for understanding the strong nuclear forces~\cite{Epelbaum:2008ga}, equation of state of the neutron stars~\cite{Akmal:1998cf,Onsi:2002qf}, nuclear structure~\cite{MYERS19661}, stellar evolution and $r$-process nucleosynthesis processes~\cite{Burbidge:1957vc,Martin:2015xql,Mumpower:2015ova,Schatz:1998zz}. However, the persistent pursuit of a precise nuclear mass remains an urgent challenge, which is benefitted from the advancements in nuclear facilities. 
Measurements on the neutron-rich side of the nuclear chart remain inaccessible, leaving most $r$-process nuclei unexplored~\cite{Ahn:2019xgh}. The neutron drip line has been experimentally confirmed only for nuclei with proton numbers up to $Z$=10. The exact boundaries of the nuclear landscape remain experimentally unknown. Consequently, our understanding of atomic properties far from stability and the ultimate limits of the nuclear landscape relies primarily on theoretical calculations~\cite{Erler2012}.

Atomic Mass Evaluation (AME) provides reliable and comprehensive experimental measurements of atomic masses~\cite{Audi:2006vh,Wapstra:2003uvz,Audi:2002rp,Audi_2012,Wang:2012eof,Huang_2017}. Theoretically, numerous models developed to pursue nuclear mass properties, have also achieved significant progress. One of the earliest approaches involved constructing phenomenological empirical models, the Bethe-Weizs$\ddot{a}$cker (BW) formula, which treats the nucleus as a charged liquid drop and approximates nuclear masses with 3 MeV accuracy~\cite{Bethe:1936zz,Weizsacker:1935bkz,Niu:2022gwo}. More accurate nuclear mass descriptions have been achieved using macroscopic–microscopic models, including the Finite Range Droplet model (FRDM2012)~\cite{Moller:2015fba}, Hartree-Fock-Bogoliubov (HFB) model ~\cite{Aboussir:1995xby,Goriely:2013xba}, and Weizs$\ddot{a}$cker-Skyrme (WS4)~\cite{Wang:2014qqa}. Covariant density functional theory (CDFT) provides a robust framework for studying nuclear structure successfully~\cite{Ring:1996qi,Vretenar:2005zz,Zhou:2016ujx}. Within this framework, the relativistic continuum Hartree–Bogoliubov (RCHB) method~\cite{Xia:2017zka}, employing the relativistic density functional PC-PK1~\cite{Zhao:2010hi} has been significantly successful in studying both stable and exotic nuclei. The deformed relativistic Hartree–Bogoliubov theory in the continuum (DRHBc) has been applied to even–even nuclei~\cite{DRHBcMassTable:2022uhi} and even-$Z$ nuclei~\cite{DRHBcMassTable:2024nvk}, yielding root-mean-square deviation (RMSD) of 1.518 MeV and 1.433 MeV, respectively. Notably, WS4 delivers the highest accuracy, with an RMSD of 0.298 MeV~\cite{Wang:2014qqa}. 
However, model accuracy deteriorates notably for heavier nuclei, especially those far from the $\beta$-stability line, with errors sometimes reaching several MeV~\cite{Utama:2017wqe}. This uncertainty presents a significant challenge for reliable predictions of exotic nuclear properties~\cite{Shang:2024dpq}.
 
In recent years, artificial intelligence (AI) has emerged as a transformative tool in nuclear physics, with machine learning (ML) algorithms significantly enhancing predictive accuracy and extrapolation capabilities. ML has been increasingly applied to various aspects of nuclear physics, achieving notable successes in predicting atomic masses~\cite{Utama:2017wqe,Lu:2024nkr,Utama:2017ytc,Li:2022ifg,Zeng:2022azv,Niu:2022gwo}, nucleon density distributions, charge radii~\cite{Ma:2020rdk,Dong:2021aqg}, charge density distributions, $\beta$ decay rates~\cite{Niu:2018trk}, $\alpha$ decay rates~\cite{Jin:2023igd,Li:2022ifg,Ma:2023ofi}, reaction cross-sections~\cite{Iwamoto:2024dtp,Mitra:2024few}, quadrupole deformations~\cite{Lin:2024okf,Lv:2024zhc}, and heavy-ion collision~\cite{Yang:2023djv,Ma:2023zfj,He:2023zin,Zhou:2023pti,Guo:2023zfk,Wang:2020tgb,Song:2021rmm,Chen:2024aom,Shou:2024uga,Gao2023,He:2023urp}. Especially, ML has been widely applied in prediction of nuclear mass, including Artificial Neural Networks (ANNs)~\cite{Li:2022ifg,Zeng:2022azv,Utama:2017ytc,Yuksel:2021nae}, Bayesian Neural Networks (BNNs)~\cite{Niu:2022gwo,Dong:2021aqg}, Convolutional Neural Networks (CNNs)\cite{Lu:2024nkr}, Light Gradient Boosting Machine (lightGBM)~\cite{Gao:2021eva}, Support Vector Machines (SVMs)~\cite{Yuksel:2024zky}, and Bayesian Gaussian Processes (BGP)~\cite{Neufcourt:2018syo}. 

Fully Connected Neural Networks (FCNNs) provide a simple yet effective framework for capturing nonlinear relationships, making them well suited for small-scale regression problems~\cite{10.1214/20-AOS2034,HORNIK1989359}. FCNNs can also effectively model the complex, nonlinear interactions between nuclear structure variables, such as proton/neutron number and binding energies. Moreover, FCNNs are particularly suitable for nuclear physics tasks, where atomic mass data is often limited. Compared to more complex architectures, FCNNs are computationally efficient and less prone to overfitting, which makes them an ideal choice for this specific task. In this study, we used FCNNs to predict nuclear masses, with the framework enhanced through the integration of two complementary strategies: direct nuclear mass prediction and residual correction between theoretical and experimental values. In addition, effective auxiliary outputs are introduced to accelerate training convergence and improve predictive accuracy by focusing on challenging-to-represent components; thereby, they enhance both training efficiency and representation quality.

Our findings show that the prediction of residual components is more effective than direct mass predictions, allowing the FCNN to address theoretical limitations with improved accuracy and interpretability. Sensitivity analysis and physical evaluations further confirm that the model captures key nuclear phenomena, such as shell corrections and pairing effects, validating its physics-driven utility. The integration of auxiliary outputs and physics-informed features highlight a strong synergy between machine learning and nuclear theory, significantly improving performance, robustness, and interpretability for precise nuclear property predictions.

The paper is organized as follows. The details of the methodology and FCNN framework are discussed in Sect.~\ref{sec:method}. Section~\ref{sec:results} provides the results and discusses the validation and extrapolation abilities of the proposed models. The summary and outlook are given in Sect.~\ref{sec:summary}.

\begin{figure*}[t!]
    \centering
    \includegraphics[width=1\textwidth]{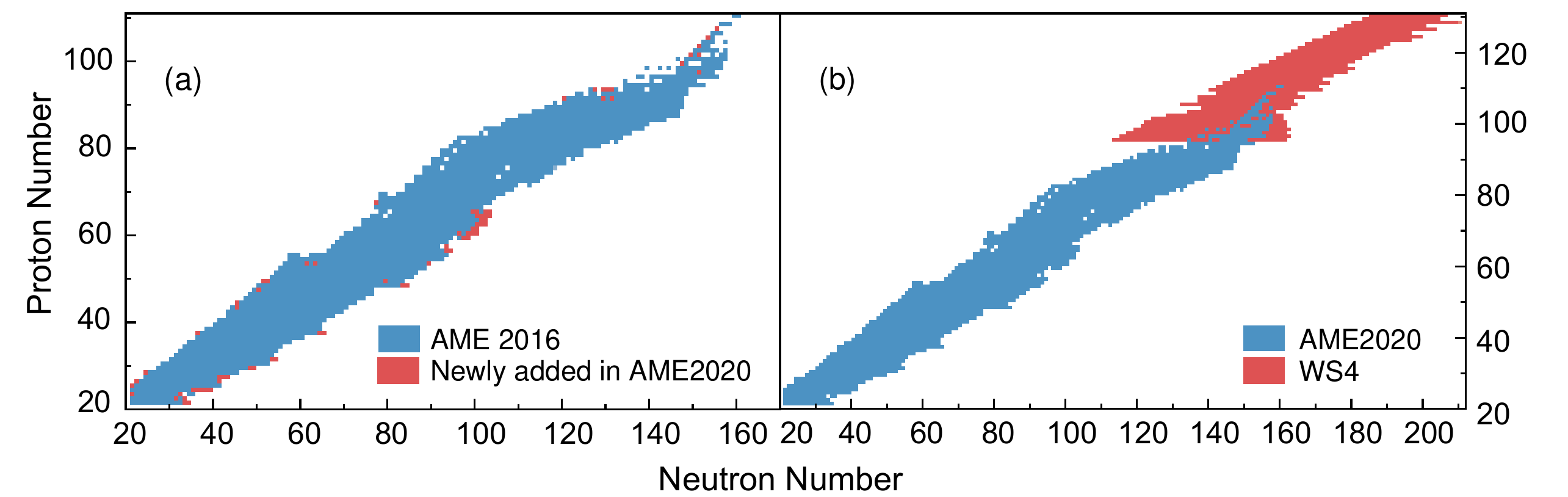}
    \caption{(a) The mass data were selected from the AME2020 and AME2016 datasets, with \(Z, N > 20\), and only the nuclei with experimental errors less than 100 keV were used.
(b) The mass data were selected from the AME2020 dataset for nuclei with \(Z, N > 20\), and only the nuclei with experimental errors less than 100 keV were used. For superheavy elements ($Z > 94$), predictions were supplemented using the WS4 model.}
    \label{fig1}
\end{figure*}

\section{Methodology and framework}\label{sec:method}

\subsection{Mass data}
\label{sec:mass}
To capture the statistical properties of nucleons, we analyze the binding energy $E$ for the nuclei with proton and neutron numbers exceeding 20, yielding 2162 and 2087 records from AME2020~\cite{Wang:2021xhn} and AME2016~\cite{AME:2016}, respectively. These datasets are divided into training and test sets for neural networks. Fig.~\ref{fig1} illustrates the distribution of the nuclei used for training and testing, highlighting the characteristics of the dataset utilized in subsequent model development. 

Distinct preprocessing strategies are applied to input features and output labels. Input features are scaled to [0, 1] using \textit{MinMaxScaler} from \textit{PyTorch}, maintaining consistency even for binary variables. Output labels are normalized to accelerate training and ensure outputs have a similar scale, improving gradient stability and prediction accuracy. Preprocessing parameters for both input features and output labels are derived solely from the training set to avoid data leakage. Predicted outputs are subsequently unscaled to recover the original nuclear binding energy.

\subsection{Model evaluations} 
The prediction accuracy and extrapolation capacity were quantified using the RMSD and mean absolute deviation (MAE), defined as:
\begin{eqnarray}
\begin{aligned}
\text{RMSD} &= \sqrt{\frac{1}{N} \sum_{i=1}^{N} (E^{\text{nn}} - E^{\text{\text{exp}}})^2}, \\
\text{MAE} &= \frac{1}{N} \sum_{i=1}^{N} \left| E^{\text{nn}} - E^{\text{\text{exp}}} \right|,
\end{aligned}
\end{eqnarray}
where $E^{\text{nn}}$ represents neural network prediction and $E^{\text{\text{exp}}}$ denotes the experimental value.

Statistical analysis of a network's predictions is insufficient to confirm its adherence to physical principles. Evaluating whether input influences align with physical laws is essential. Sensitivity analysis quantifies the impact of each input feature on prediction quality. The normalized gradient method is employed to access the contributions of input features. This approach computes the absolute gradient of the neural network with respect to each feature and normalizes the values to determine relative importance. Gradients are derived from all available data during training and evaluation. The sensitivity score and normalized sensitivity score are defined as:
% Gradient Absolute Value Method
\begin{eqnarray}
\begin{aligned}
&S_j = \frac{1}{N} \sum_{i=1}^N \left| \frac{\partial y_i}{\partial x_j} \right|, \\
&S_j^{\text{norm}} = \frac{S_j}{\sum_{j=1}^M S_j},
\end{aligned}
\end{eqnarray}
where $S_j$ denotes the sensitivity score of the $j$-th feature, $N$ is the number of samples, $y_i$ is the model output for the $i$-th sample, and $x_j$ is the $j$-th input feature. $S_j^{\text{norm}}$ represents normalized sensitivity score, with $M$ being the total number of features.

The models selected for subsequent experiments are evaluated according to three key criteria. The model must perform well on the training and test sets, with closely aligned results, ensuring good generalization. Input features must align with the underlying physical background to evaluate their impact on the output. For the extrapolation task concerning the superheavy nuclei far from the $\beta$-stability line, the results must not only satisfy numerical accuracy from the first criterion but also reflect physical phenomena driven by input features in the training set, capturing embedded physical information.

\subsection{Input features and theoretical framework}
\label{sec:inputfeatures}
We extract the input features using a macroscopic-microscopic approach that combines two models: the improved semi-classical liquid drop model and the shell model. This approach incorporates the physical background into the input features.

The improved semi-classical liquid drop model describes the nucleus as a positively charged, incompressible droplet with nucleons behaving like molecules. The binding energy in the improved liquid drop model can be expressed as~\cite{Wang:2010uk}
\begin{eqnarray}
E_{\text{LD}}(A,Z) = E_v + E_s + E_C + E_{\text{sym}} + E_{\text{pair}},
\label{eq:E_LD}
\end{eqnarray}
where the right-hand side represents the following terms:
\begin{eqnarray}
\label{eq:ELDterms}
E_v &=& a_v A, \nonumber\\
E_s &=& a_s A^{2/3}, \nonumber\\
E_C &=& a_c \frac{Z^2}{A^{1/3}} \left( 1 - Z^{-2/3} \right), \\
E_{\text{sym}} &=& a_{\text{sym}} I^2 A,  \nonumber\\
E_{\text{pair}} &=& a_{\text{pair}} A^{-1/3} \delta_{np}.\nonumber
\end{eqnarray}
These terms correspond to volume energy, surface energy, Coulomb energy, asymmetry energy, and pairing energy, respectively. Here, \(a_i\) are the corresponding coefficients, \(I = (N - Z)/A\) is the isospin asymmetry, and \(\delta_{np}\) encodes the parity of the nucleus (whether it is odd or even in neutron and proton numbers).

The shell model provides a fundamental framework for understanding nuclear stability, particularly near magic numbers~\cite{Duflo:1995ep}. To quantify this, the valence proton number \(V_p\) and valence neutron number \(V_n\) are introduced as input features. These are defined as
\begin{eqnarray}
\begin{aligned}
V_p &= Z - Z_{\text{core}}, \\
V_n &= N - N_{\text{core}},
\end{aligned}
\end{eqnarray}
where \(Z_{\text{core}}\) and \(N_{\text{core}}\) denote the nearest magic numbers smaller than the given $Z$ and $N$. In this study, the magic number adopted is ($Z=$ 20, 28, 50, 82, 114, 120 and $N=$ 20, 28, 50, 82, 126, 184, 198), based on the reference ~\cite{Sorlin:2008jg,Bender:2003jk,Meng:2005jv}.

The general input features derived from the above descriptions are listed in Tab.~\ref{tab1} and expressed as follows. \(Z\) and \(A\) characterize the nuclear species, with the binding energy depending on both. \(A\) represents the volume energy, while \(A^{2/3}\) reflects the surface energy, and $E_C/a_c$ represents the Coulomb energy. The symmetry energy is described by \(I^2 A\) and \(|I|\). Finally, \(A^{-1/3}\) and \(\delta_{np}\) reflect the contribution of the pairing term.
Additionally, for the shell correction, we introduce \(V_p\) and \(V_n\), representing the valence protons and neutrons, respectively.

\begin{table*}[htpb]
    \centering
    \caption{Input and output features in Method I, II, and III.}
    \label{tab1}
    \renewcommand{\arraystretch}{1.5} % Adjust row height for readability
    \begin{tabular}{c|cc|c}
        \hline\hline
        Methods &    \multicolumn{2}{c|}{ Input Features} & Output Features \\ \hline
        I
            &$Z,A,A^{2/3},E_C/a_c,I^2 A,|I|,A^{-1/3},V_p,V_n,\delta_{np}$,~~~~~& \(A^{-1/2}\) & \( E^{\text{exp}} \) \\ 
        II
            &$Z,A,A^{2/3},E_C/a_c,I^2 A,|I|,A^{-1/3},V_p,V_n,\delta_{np}$,~~~~~& \(g_1 A^{1/3},g_2A^{-1/3}\) & \( E_\text{LD},E^{\text{exp}},E^{\text{exp}} - E_\text{LD} \)  \\ 
        III
            &$Z,A,A^{2/3},E_C/a_c,I^2 A,|I|,A^{-1/3},V_p,V_n,\delta_{np}$,~~~~~& \( \prod_{k \in \{2, 4, 6\}} b_k \beta_k^2 \) & \( E_\text{LD},E^{\text{exp}},E^{\text{exp}} - E_\text{LD} \)  \\
            \hline\hline
    \end{tabular}
\end{table*}

In addition to the aforementioned general input features, we explore three distinct methods to incorporate the effects of deformation, microscopic corrections, and multi-output, as outlined in Tab.~\ref{tab1}.
\subsubsection{Method I}
\label{sec:method1}
In this method, we consider versions of the liquid droplet model where the pairing term takes the form \( E_{\text{pair}} = a_{\text{pair}} A^{-1/2} \delta_{np} \)~\cite{greiner1996nuclear}. To incorporate this, we introduce an additional input feature \( A^{-1/2} \) to compete with the \( A^{-1/3} \) term discussed above. We encode \( \delta_{np} \) as a two-dimension vector using the following definition:
\begin{equation}
\delta_{np} =
\begin{cases} 
(1, 0), & \text{if $N$ and $Z$ are even} \\
(0, 1), & \text{if $N$ and $Z$ are odd} \\
(0, 0), & \text{other}.\\
\end{cases}
\label{eq:deltanp1}
\end{equation}
For the output features, we adopt the experimental binding energy $E^{\text{exp}}$ exclusively, as listed in Tab.~\ref{tab1} (Method I).

\subsubsection{Method II}

To enhance prediction accuracy, we developed an alternative approach, Method II, which incorporates revised input features, output features, and network architecture for comparison with Method I. The total binding energy of a nucleus, which includes deformation and Strutinsky shell corrections, \( E'(A, Z, \beta) \), is expressed as
\begin{equation}
\label{eq:mac-mic}
E(A, Z, \beta) = E_{\text{LD}}(A, Z) \prod_{k \in \{2, 4, 6\}} \left( 1 + b_k {\beta_k}^2 \right) + E'(A, Z, \beta),
\end{equation}
where the curvature of the parabola for a given \( \beta_k (k=2,4,6)\) deformation is approximately described by the empirical formula~\cite{Wang:2010uk}
\begin{equation}
\label{eq:curvature_parabola}
b_k = (\frac{k}{2}){g_1 A^{1/3}} + {(\frac{k}{2})^2}{g_2 A^{-1/3}}.
\end{equation}
Given the limited data available for \( \beta_k \) and its dependence on \( A \) and \( Z \), its effects are captured indirectly through these variables. Moreover, the contribution of \( b_k \) is modeled by decomposing it into two components for the network input: \( g_1 A^{1/3} \) and \( g_2 A^{-1/3} \). The macro-micro model, as described in Eq.~(\ref{eq:mac-mic}), extends the liquid-drop model by incorporating both deformation and microscopic corrections.
Neural networks typically perform better when the output labels are strongly correlated~\cite{song2022learning,jia2020residual}. 

In this method, we introduce two additional output labels to constrain the primary target, the difference between experimental values and those of the liquid-drop model. This constraint enhances the loss function and improves the accuracy of prediction. Theoretical binding energies, which are highly accurate, are excluded as output labels to avoid redundancy. However, predicting experimental values introduce a challenge, as it risks conflating microscopic and macroscopic corrections, which will potentially increase the error.
To address this issue, we isolate the microscopic component by subtracting the macroscopic part (including the droplet and deformation terms) from the total binding energy. This approach allows the model to focus specifically on capturing the intricate microscopic features.

To better leverage experimental and theoretical data, we propose a multi-output approach with three output labels: the experimental binding energy (\( E^{\text{exp}} \)), the liquid-drop model binding energy (\( E_\text{LD} \)), and their difference (\( E^{\text{exp}} - E_\text{LD} \)), as listed in Tab.~\ref{tab1} (Method II). This strategy results in significantly improved accuracy over Method I.
The parameters used to calculate \( E_\text{LD} \) in this method are taken from Ref.~\cite{Wang:2010uk}.

\subsubsection{Method III}
\label{sec:method3}
In this method, we adopt a strategy similar to Method II but with a few minor adjustments, as listed in Tab.~\ref{tab1} (Method III). Specifically, the Coulomb term differs from that in Eq.~(\ref{eq:ELDterms}) and takes the following form, as used in the WS4 model~\cite{Wang:2014qqa}.
\begin{eqnarray}
\label{eq:EC2}
E_C &=& a_c \frac{Z^2}{A^{1/3}} \left( 1 - 0.76 Z^{-2/3} \right).
\end{eqnarray}
Next, the parameter \( \delta_{np} \) is encoded as follows
\begin{equation}
\delta_{np} =
\begin{cases} 
2 - |I|, & \text{if $N$ and $Z$ are both even} \\[5pt]
|I|, & \text{if $N$ and $Z$ are both odd} \\[5pt]
1 - |I|, & \text{if $N$ is even, $Z$ is odd, and $N > Z$} \\[5pt]
1 - |I|, & \text{if $N$ is odd, $Z$ is even, and $N < Z$} \\[5pt]
1, & \text{if $N$ is even, $Z$ is odd, and $N < Z$} \\[5pt]
1, & \text{if $N$ is odd, $Z$ is even, and $N > Z$}.
\end{cases}
\label{eq:deltanp2}
\end{equation}
For the deformation contribution, we replace the terms \( g_1 A^{1/3} \) and \( g_2 A^{-1/3} \) in Eqs.~(\ref{eq:mac-mic}) and~(\ref{eq:curvature_parabola}) with the summation form defined as: 
\[
{def} = \prod_{k \in \{2, 4, 6\}} b_k \beta_k^2,
\]
which provides a more accurate description of the nuclear shape. Finally, the parameters used to calculate \( E_\text{LD} \) in this method and $\beta_k$ for the given nucleus are taken from Ref.~\cite{Wang:2014qqa}.
Through these adjustments, this method achieves improved accuracy while maintaining a framework similar to that of Method II.

\subsection{Network architecture}
\label{sec:network}
\begin{figure}[htpb]
    \centering    \includegraphics[width=0.485\textwidth]{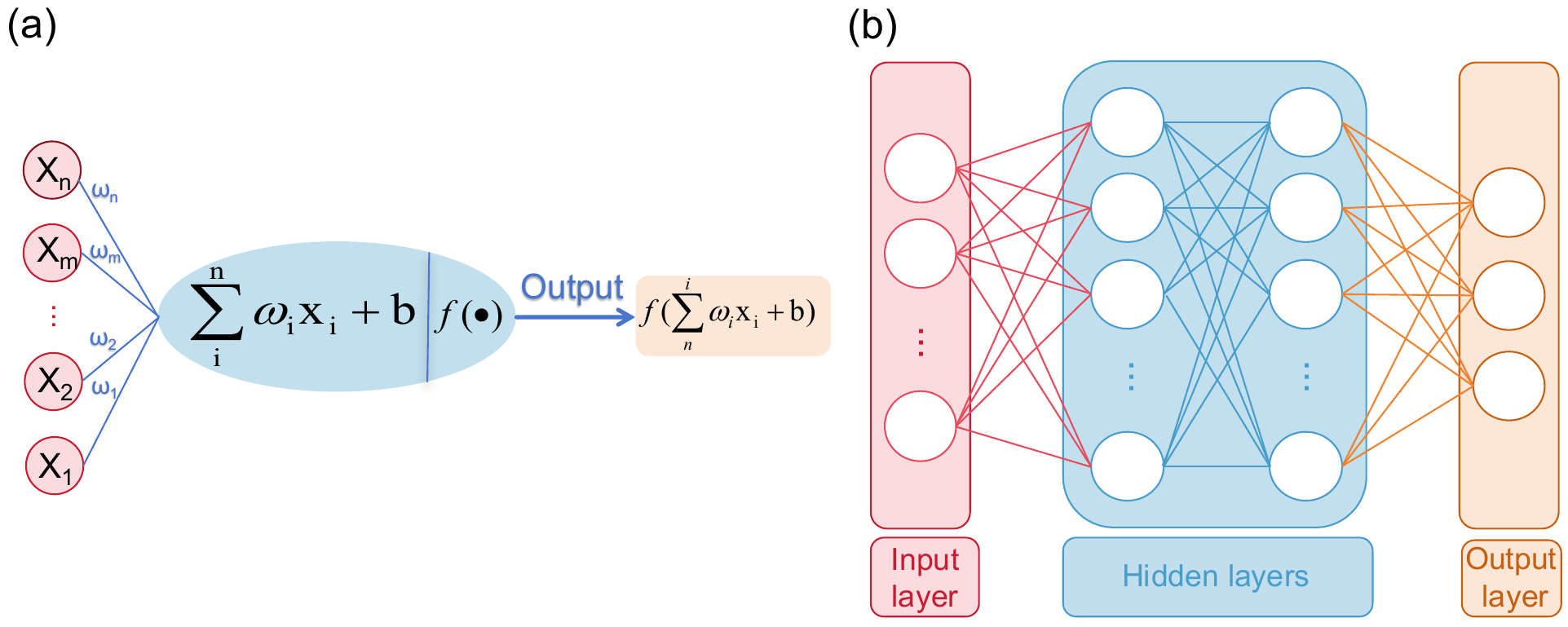}
    \label{fig2}
    \caption{(a) The single-layer perceptron consists of an input layer, weights, a bias, and an output layer, performing a nonlinear mapping via a linear transformation and activation function $f$. (b) The FCNN is composed of multiple SLPs, with numerous weights and bias terms.}
    \label{fig2}
\end{figure}
An FCNN with two hidden layers was used for optimal feature representation and prediction. The first layer, 80–120 neurons, captured input features, while the second, with 40–60 neurons, refined these features. Our model's network structure selection criterion is that if the training set performance is similar, we choose the one that performs better on the test set. If the test set performance is similar, we choose the structure with closer training and test set performances. If both the training and test set performances are close, we select the structure that has a faster training time and more stable convergence.
This design balances model complexity and efficiency, ensuring adaptability to diverse samples. Preliminary experiments revealed that the prediction of binding energy was slightly improved by adding 1-5 layers but the training time and overfitting risks were also increased. Similarly, increasing neurons per layer up to 500 enhanced the performance but posed similar challenges. In contrast, the overly simple network led to slower training and poor fitting. These findings highlight the need to balance network complexity and generalization for efficient training and robust performance. The mechanisms of the single-layer perceptron (SLP) and the neural network architecture are illustrated in Fig.~\ref{fig2}.

We selected the GELU activation function in FCNN for its smooth transition between positive and negative values, enhancing gradient flow and enabling neurons to fully learn input features with physical significance. GELU outperforms ReLU and Leaky-ReLU in mitigating gradient vanishing issues and is more accurate and efficient for regression problems than Sigmoid and Tanh~\cite{hendrycks2016gaussian,lee2023gelu}.

We avoided decay mechanisms like Adam’s learning rate (LR) decay or cosine decay, which can cause the optimizer to get stuck in local minima. Instead, we manually adjusted the learning rate based on loss behavior. Initially, a larger learning rate was used. After each reduction in learning rate, we record the value of the loss function where the next oscillation begins. The experiment is repeated, selecting the highest loss value and then reducing the learning rate slightly beyond that point. This process continues until the model becomes difficult to converge, allowing us to establish a relationship between the learning rate and the loss function. Finally, multiple training runs are performed to ensure that the chosen learning rate guarantees stable convergence.

The Huber loss function is used for its robustness to outliers. Unlike squared loss, it remains stable with large discrepancies between theoretical and experimental values, preventing overfitting~\cite{Huberloss}. Its differentiability when values match makes it effective for regression tasks, defined as follows
\begin{eqnarray}
\begin{aligned}
\begin{small}
L_\text{Huber} = 
\begin{cases} 
\frac{1}{2}(E^{\text{nn}}-E^{\text{\text{exp}}})^2, |E^{\text{nn}} - E^{\text{\text{exp}}}| \leq \delta, \\
\delta \cdot (|E^{\text{nn}}-E^{\text{\text{exp}}}|-\frac{1}{2} \delta), |E^{\text{nn}} - E^{\text{\text{exp}}}| > \delta, 
\end{cases}
\end{small}
\end{aligned}
\end{eqnarray}
with $\delta$ = 1, $E^{\text{nn}}$ is FCNN's prediction as a function of model weights and bias, and $E^{\text{exp}}$ is the experimental value. 

To improve the network's generalization and ensure consistent performance across training and test sets, we introduce $L_2$ regularization. This technique adds the sum of squared model parameters ($L^2$ norm) as a penalty term to the loss function, keeping model weights small. For our regression task, $L_2$ regularization is applied via weight decay in the Adam optimizer. The modified loss function is
\begin{eqnarray}
\begin{aligned}
L_{\text{reg}}(w) = L(w) + \frac{\lambda}{2} \|w\|_2^2,
\end{aligned}
\end{eqnarray}
with $\lambda$ controls the regularization strength. We started with a relatively large regularization coefficient of 0.001, which caused the model to converge too quickly. By gradually reducing the regularization strength, we ensured that the model would converge just right within a sufficient number of epochs in the training set.

Finally, we clarify two key differences in the strategies for improving generalization ability across the three methods. In Methods II and III, the learning rate doesn't need to be divided into multiple stages (or even at all) to achieve improved performance on the training set. The test set performance improves initially but then starts to decline, which may be due to the auxiliary outputs shortening the model's training path. Using $L_2$ regularization to improve generalization is applied only in Method I, as its longer training path makes it challenging to monitor the relationship between the training and validation set performance in the same way as with the other two methods.

\section{Results and discussions} \label{sec:results}

We study three different cases across three FCNN methods. In Case A, models were trained on AME2020 mass data and split 8:2 into training and test sets. All the methods demonstrated strong performance, accurately representing nuclear masses. To further examine the models, a sensitivity analysis was conducted to assess the impact of input features on outputs, providing insights into the neural networks and making the ``black box" appear slightly more transparent. In Case B, we assessed the models' extrapolation abilities by training on AME2016 data and predicting newly added data in AME2020. Through optimization, the neural networks achieved reasonable extrapolation capability. In Case C, building on the superior performance, extrapolation capability, and efficiency of the second method, we expanded Case A's dataset to include WS4 model results, which were absent in the AME2020 data. Following the same training and validation protocol,  we verified the predicted outputs by calculating physical quantities to ensure consistency with governing physical laws.
\begin{figure}[htbp]
    \centering    
    \includegraphics[width=0.4\textwidth]{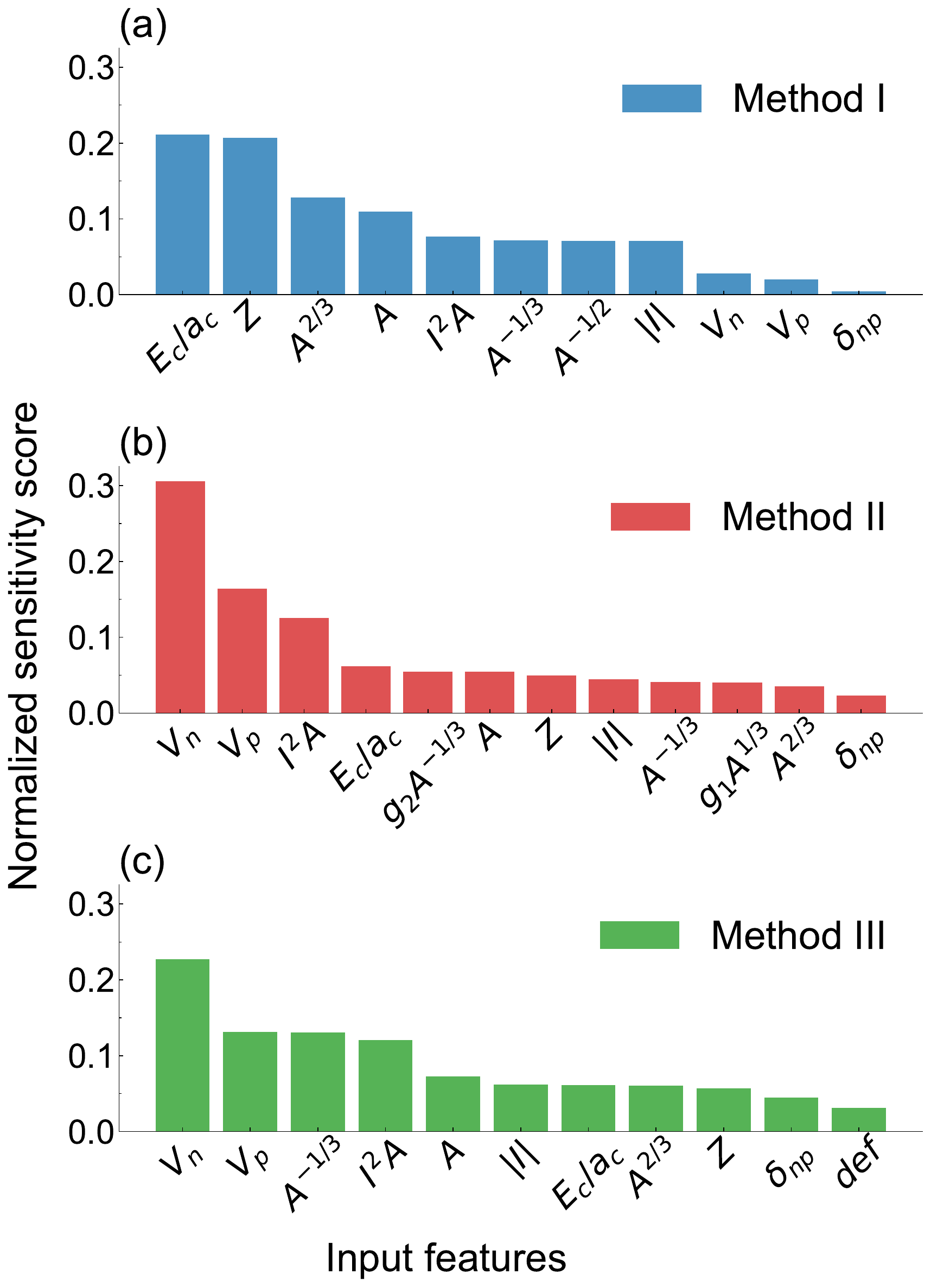}
    \label{figsen}
    \caption{Comparison of normalized sensitivity analysis in different input features among three Methods.}
    \label{fig3}
\end{figure}

\subsection{Case A: performance in the AME2020}
\label{sec:caseA}
Figure~\ref{fig3} shows the sensitivity analysis in three distinct methods in Case A. As observed, macroscopic features dominate in Method I, while microscopic features derived from the shell model exert a stronger influence in Method II and III. Variations in $\delta_{np}$ show less impact on all the results, as Method I focuses on the overall binding energy predictions, and Method II targets the more challenging region-dominated nuclear shell effects. Although $\delta_{np}$ are highly correlated with $A$ and $Z$, and thus partially embedded in other features, their significance remains indispensable. To validate this, we conducted ablation tests by removing specific features and analyzing the change in RMSD, using the same optimization strategies as before. For Method I, the training and test RMSD are 0.756 MeV and 0.852 MeV, respectively, while Method II's training and test RMSD are 0.181 MeV and 0.223 MeV, respectively. Method III's training and test RMSD are 0.172 MeV and 0.196 MeV. In Method III, variations in $def$ show less impact on predictions, and we also conducted an ablation test by removing $def$ while using the same optimization strategies as before. Finally, Method III's training and test RMSD are 0.069 MeV and 0.154 MeV, respectively. These results indicate that Method I relies more on $\delta_{np}$ but all the methods require them as inputs to achieve optimal performance.

\begin{table}[h!]
    \centering
    \caption{Loss function vs. learning rate for Case A, B, and C in three Methods. }
    \label{tab3}
    \begin{tabular}{c|c|c|c}
        \toprule
        Case & Method & Loss function ($\times$ $10^{-6}$) & Learning rate ($\times$ $10^{-6}$) \\  \hline
        \multirow{6}{*}{A} & \multirow{5}{*}{I} & $\ge$ 40 & 3000  \\ 
                           &                         & [10, 40) & 500 \\ 
                           &                         & [7.5, 10) & 50  \\ 
                           &                         & [6, 7.5) & 20 \\ 
                           &                         & $<$ 6 & 3  \\ \cline{2-4}
                           & II/III                      & $\ge$ 0 & 1000 \\ \hline
        \multirow{8}{*}{B} & \multirow{8}{*}{I} & $\ge$ 50 & 3000  \\ 
                           &                         & [25, 50) & 1000 \\ 
                           &                         & [20, 25) & 500 \\ 
                           &                         & [12, 20) & 100 \\
                           &                         & [10, 12) & 20 \\
                           &                         & [8, 10) & 10 \\
                           &                         & $<$ 8 & 3  \\ \cline{2-4}
                           & II/III                      & $\ge$ 0 & 100 \\ \hline
        \multirow{3}{*}{C} & \multirow{3}{*}{III} & $\ge$ 50 & 3000  \\ 
                           &                         & [20, 50) & 500 \\ 
                           &                         & $<$ 20 & 100  \\ \hline\hline
    \end{tabular}
\end{table}

\begin{table}[h!]
    \centering
    \caption{Training and test set errors for Case A, B, and C in three Methods.}
    \label{tab4}
    \begin{tabular}{ccccc}
        \toprule
        Case & Method & Datasets & RMSD (MeV) & MAE (MeV) \\ \hline
        \multirow{9}{*}{A} & \multirow{3}{*}{I}  & Training & 0.135 & 0.099 \\ 
                           &                    & Test  & 0.204  & 0.141 \\% \cline{2-5}
                           &                    & All  & 0.151  & 0.107 \\
                           & \multirow{3}{*}{II} & Training & 0.087 & 0.064  \\ 
                           &                    & Test  & 0.143 & 0.099 \\ 
                           &                    & All  & 0.101  & 0.071 \\
                           & \multirow{3}{*}{III} & Training & 0.052 &  0.039 \\ 
                           &                    & Test  & 0.122 & 0.087 \\ 
                           &                    & All  & 0.071  & 0.049 \\ \hline
        \multirow{9}{*}{B} & \multirow{3}{*}{I}  & Training & 0.195 & 0.145 \\ 
                           &                    & Test  & 0.416 & 0.283 \\ %\cline{2-5}
                           &                    & All  & 0.207 & 0.150 \\
                           & \multirow{3}{*}{II} & Training & 0.159 & 0.122 \\ 
                           &                    & Test  & 0.237 & 0.173 \\
                           &                    & All  & 0.162 & 0.124 \\
                           & \multirow{3}{*}{III} & Training & 0.125 & 0.096 \\ 
                           &                    & Test  & 0.191 & 0.144 \\
                           &                    & All  & 0.128 & 0.097 \\\hline
        \multirow{3}{*}{C} & \multirow{3}{*}{III}  & Training & 0.116 &  0.082 \\ 
                           &                    & Test  & 0.161 & 0.107 \\
                           &                    & All  & 0.126  & 0.087 \\\hline \hline

    \end{tabular}
\end{table}

In Case A, we select the mass data from AME2020 with absolute errors less than 100 keV and $Z, N > 20$. To ensure reproducibility and minimize the impact of data splitting, we applied a specific data shuffling strategy with \textit{random\_state = 42}, a parameter for pseudo-random number generation in \textit{Scikit-learn}. All the methods used the same neural network architecture, consisting of two hidden layers with 80 and 40 neurons, respectively, to compare their extrapolation capabilities. In Method I, we observed that achieving strong performance on both the training and test sets required a significantly prolonged training process, with a risk of local minima. To address this, a dynamic learning rate was employed, adjusting according to the loss function value, as shown in Tab.~\ref{tab3} Case A. Meanwhile, an ${L_2}$ regularization coefficient of 5 $\times$ $10^{-8}$ was applied to ensure that the training set would just converge at 3 million epochs. Unlike Method II, this approach did not require pre-experimentation for early stopping. Training was halted after 2.5 million epochs, when performance stabilized within 2–3 million epochs. The resulting RMSD for the training set was 0.135 MeV, with an MAE of 0.099 MeV, as shown in Tab.~\ref{tab4} Case A. For the test set, their values were 0.204 and 0.141 MeV, indicating strong generalization. We also list the resulting values calculated from all datasets.

\begin{figure}[htbp]
    \centering\includegraphics[width=0.85\linewidth]{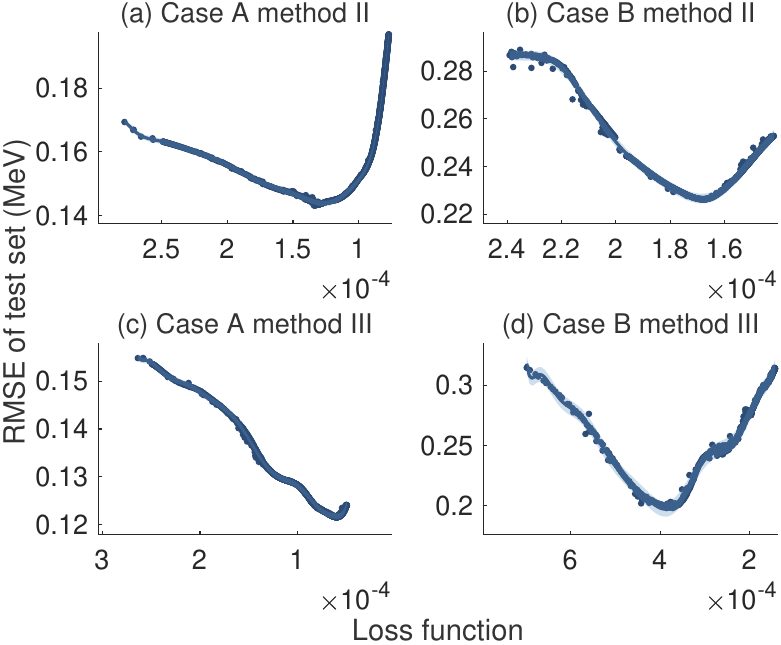}
    \caption{Variation of the RMSD of the test set with the values of the loss function in Case A and B using Method II and III. (a) presents the improved extrapolation within an optimal loss range from $1 \times 10^{-4}$ to $1.5 \times 10^{-4}$, terminating at $1.28 \times 10^{-4}$. (b) presents the enhanced extrapolation within $1.6 \times 10^{-4}$ to $1.8 \times 10^{-4}$, terminating at $1.68 \times 10^{-4}$. (c) presents the enhanced extrapolation within $5 \times 10^{-5}$ to $7 \times 10^{-5}$, terminating at $5.97 \times 10^{-5}$. (d) presents the enhanced extrapolation within $3 \times 10^{-4}$ to $5 \times 10^{-4}$, terminating at $3.80 \times 10^{-4}$. Points represent test set performance for different loss function stopping values, the curve is a polynomial fit, and the shaded area represents fitting error.} 
    \label{fig4}
\end{figure}

Method II and III adopt the same network structure selected by preliminary experiments, training, and division of test set as Method I. An early stopping strategy, shown in Fig.~\ref{fig4} (a) and (c) for Method II and III, was employed to terminate the training process based on different loss thresholds for the two methods. For Method II, training was halted when the loss reached a threshold of $1.28 \times 10^{-4}$ after nearly 30000 epochs, as further reduction did not consistently improve the performance of the test set. For Method III, a threshold of $5.97 \times 10^{-5}$ was applied, leading to termination after approximately 145000 epochs.

Unlike Method I, both Method II and III utilized a static learning rate of 0.001, achieving satisfactory test performance efficiently while conserving computational resources. The comparison of the three methods is presented in Tab.~\ref{tab3}.

\begin{figure}[htbp]
\centering
\includegraphics[width=0.75\linewidth]{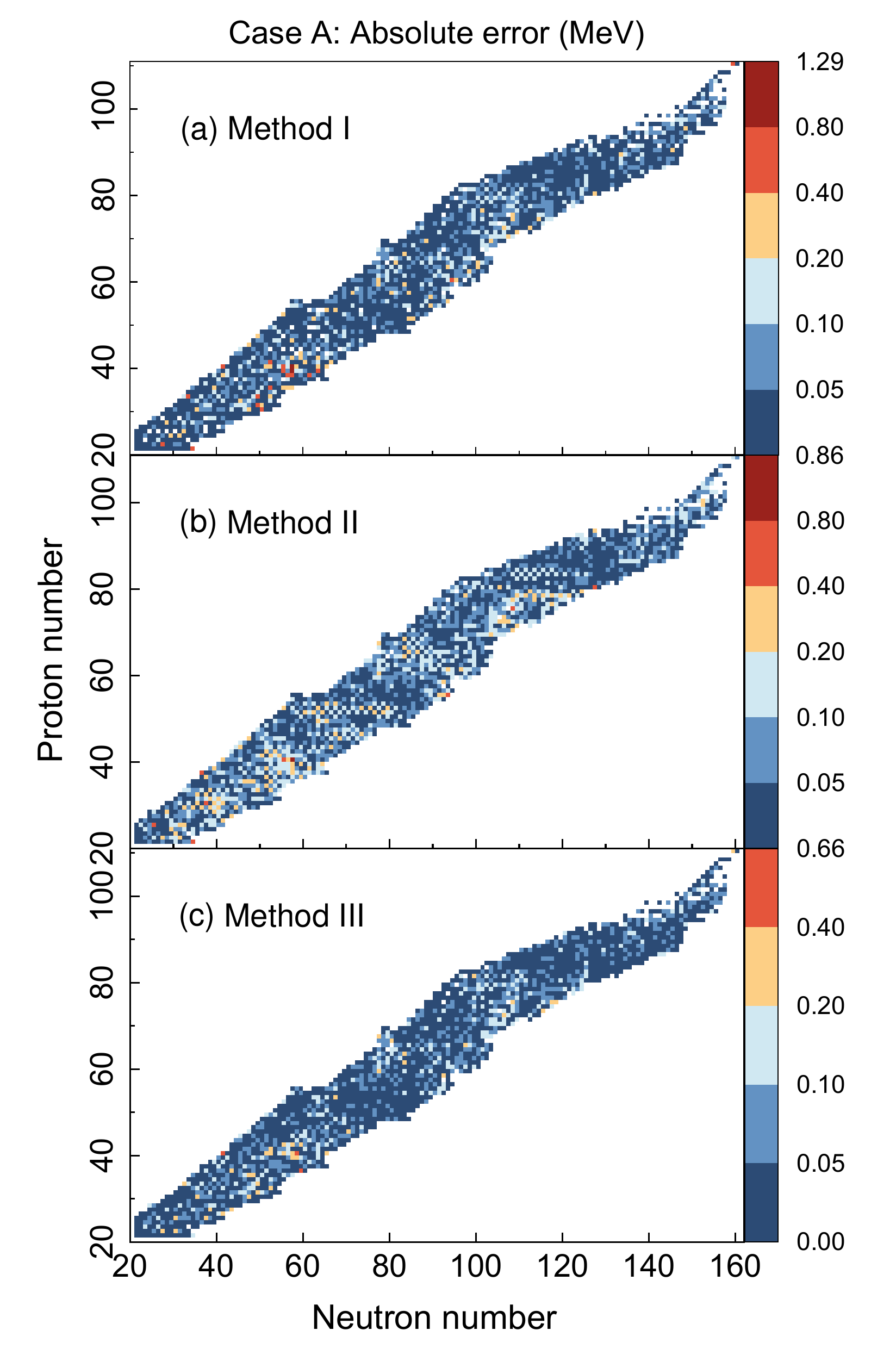}
    \caption{Prediction errors of three Methods on the AME2020 dataset, with the majority of nuclei exhibiting errors below 100 keV.
    }
    \label{fig5}
\end{figure}

After training, Method II demonstrated superior accuracy, with RMSD and MAE values of 0.087 MeV and 0.064 MeV for the training set, and 0.143 MeV and 0.099 MeV for the test set. Meanwhile, Method III further enhanced the model's accuracy, achieving RMSD and MAE values of 0.052 MeV and 0.039 MeV for the training set, and 0.122 MeV and 0.087 MeV for the test set. Detailed results for all three methods are presented in Tab.~\ref{tab4}. Figure~\ref{fig5} visualizes the improved efficiency and accuracy of both Method II (b) and Method III (c) compared to Method I (a) of the prediction errors on the AME2020 dataset and also illustrates the improved accuracy and generalization capability of Method III compared to the other two methods. 

\begin{figure}[htbp]
\centering
\includegraphics[width=0.75\linewidth]{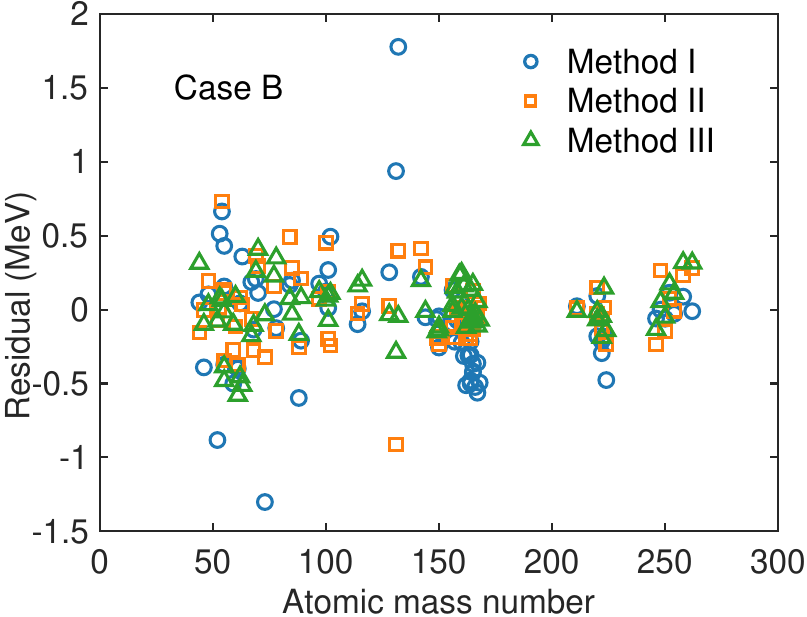}
    \caption{Comparison of the residual in three Methods in newly added data from AME2020.
    }
    \label{fig6}
\end{figure}
\subsection{Case B: prediction of newly added data in AME2020 based on AME2016}
\label{sec:caseB}
\begin{figure*}[t]
    \centering
    \includegraphics[width=0.85\textwidth]{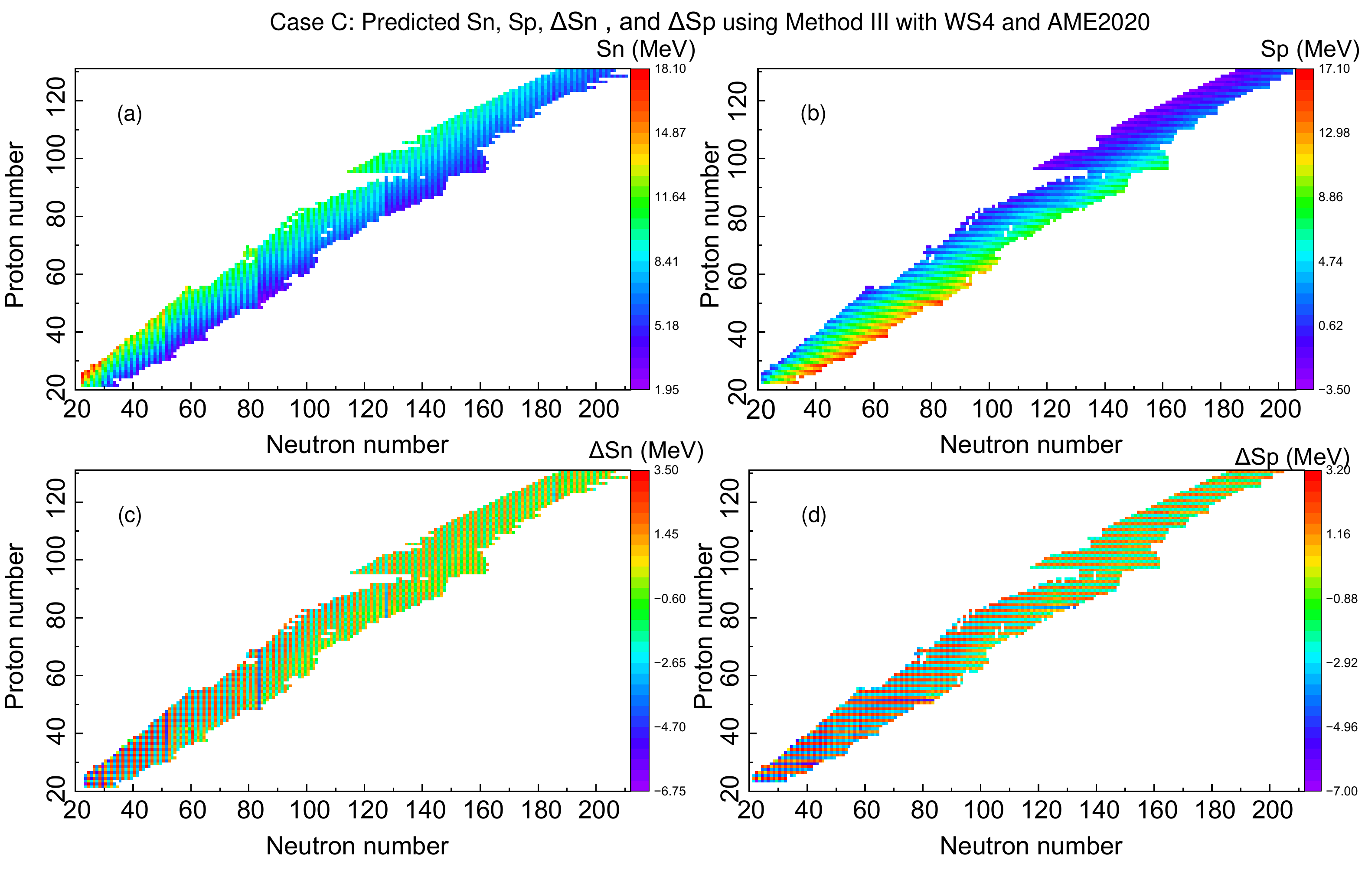}
    \caption {Predicted neutron ($S_n$) and proton ($S_p$) separation energy and their rate of changes $\Delta S_n$ and $\Delta S_p$.}
    \label{fig7}
\end{figure*}
In case B, as shown in Fig.~\ref{fig1} (a), we use the entire dataset from AME2016 as the training set, while the data included in AME2020 but not in AME2016 serve as the test set. Training on AME2016 and predicting the data newly added in AME2020 could verify the model's extrapolation capacity. It also provides a basis for a neural network to predict the data not included in AME2020, albeit within a limited extrapolation range. In Method I, an ${L_2}$ regularization coefficient of 5 $\times$ $10^{-7}$ was applied, using the same loss and activation functions as in prior Case A. A dynamic learning rate strategy was employed, with specific learning rate settings detailed in Tab.~\ref{tab3}. After 3 million epochs, the model achieved an RMSD of 0.195 MeV and an MAE of 0.145 MeV on the training set, while 0.416 MeV and 0.283 MeV on the test set, as shown in Tab.~\ref{tab4}.

In Method II and III, the network architecture features two hidden layers with 120 neurons and 60 neurons, respectively, which were selected by preliminary experiments. The loss and activation functions remained consistent with previous Case A. Training was terminated via early stopping at a threshold loss value of $1.68 \times 10^{-4}$ for Method II and $3.80 \times 10^{-4}$ for Method III, as shown in Fig.~\ref{fig4} (b) and (d).

Unlike Method I, which applied ${L_2}$ regularization and a dynamic learning rate, Method II and III used neither, as its larger architecture provided a more stable training process without requiring additional techniques to prevent overfitting. This approach simplified the optimization process by maintaining a constant learning rate throughout. Method II achieved an RMSD of 0.159 MeV and an MAE of 0.122 MeV on the training set, and an RMSD of 0.237 MeV and an MAE of 0.173 MeV on the test set, as shown in Tab.~\ref{tab4}. Method III achieved an RMSD of 0.125 MeV and an MAE of 0.096 MeV on the training set, while 0.191 MeV and 0.144 MeV on the test set, as shown in Tab.~\ref{tab4}.

Figure~\ref{fig6} shows the performance comparison of three Methods. It shows that Method II significantly outperforms Method I for most data points, with the RMSD of Method II reducing values to nearly $40\%$. Method III significantly outperforms Method II, with the RMSD of Method III reducing values to nearly $20\%$, highlighting its superior extrapolation capacity.

\subsection{Case C: nucleon pairing correlation, magic number effects, and extrapolations in isotopic chains}
\label{sec:caseC}
\begin{figure*}[t]
    \centering
    \includegraphics[width=0.75\textwidth]{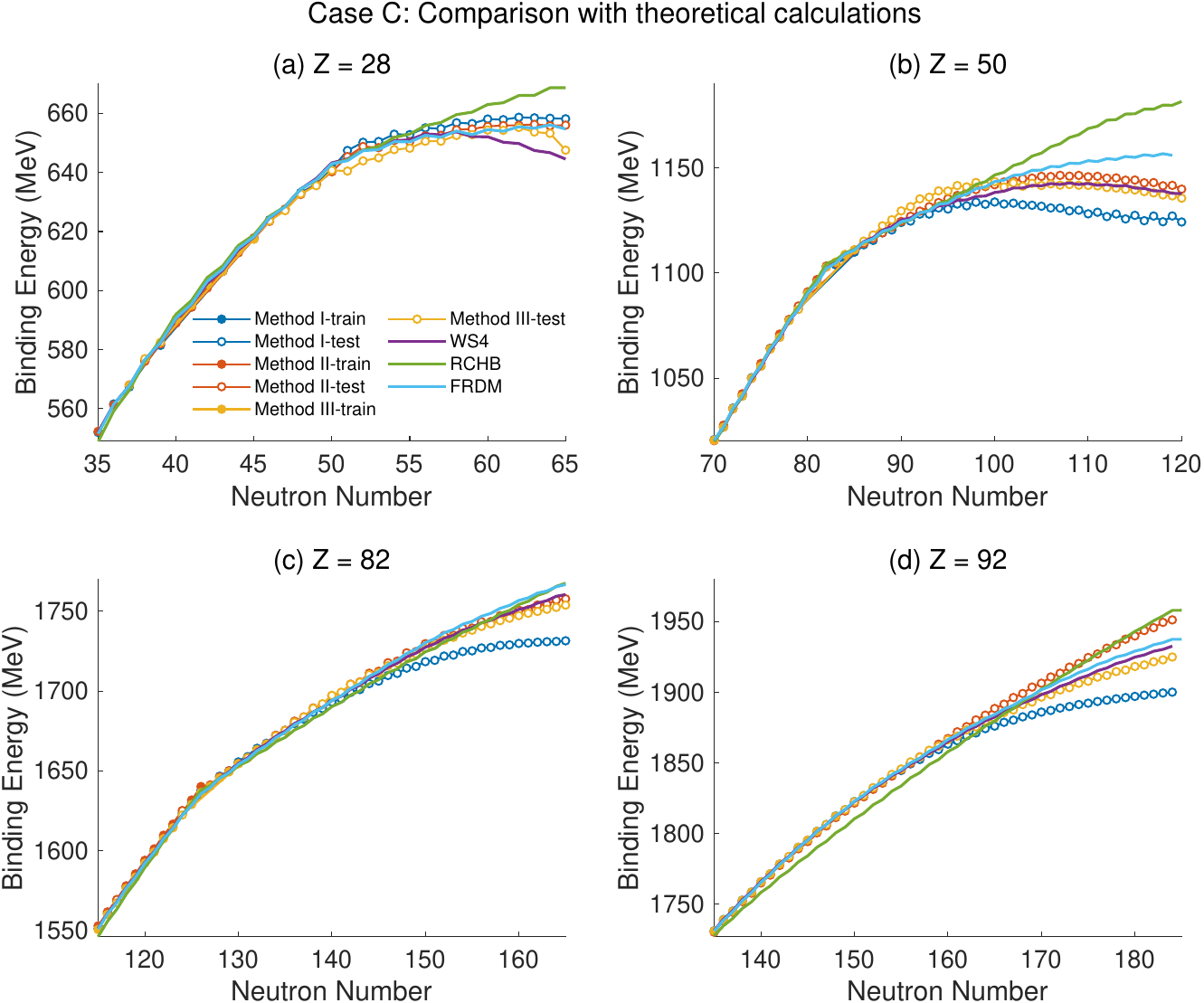}
    \caption {Comparison of extrapolation in different mass models for (a) $Z=28$, (b) $Z=50$, (c) $Z=82$, and (d) $Z=92$ isotopes in Method I, II, and III. Lines denote WS4~\cite{Wang:2014qqa}, RCHB~\cite{Xia:2017zka}, and FRDM2012~\cite{Moller:2015fba} mass models for comparisons.}
    \label{fig8}
\end{figure*}
To further evaluate Method III, we expand the training data from Case A to examine its ability to capture the physics information embedded in the inputs and reproduce certain physics phenomena. Using AME2020 data, we incorporated theoretical values from the WS4 model as auxiliary inputs to assist in the extrapolation process in the Case C study, as shown in Fig.~\ref{fig1} (b). We limited the data selection since we couldn’t assess the error of the theoretical values. We chose only the WS4 data with $Z>$ 94, as experimental data for $Z<$ 95 was sufficient for validating the magic number. We also found that the error in the theoretical values was smaller when the mass excess was lower. Therefore, we selected the isotope with the smallest theoretical mass excess for each $Z$ and expanded the range in both directions, with the mass excess ranging from 50 to 400 MeV. In the extrapolated region, corrections accounting for the proton and neutron magic number were applied. The training process adhered to the same methodology as Method III in Case A, with the selected learning rate distribution in Tab.~\ref{tab3}, and the early stopping criterion is adopted. The final RMSD values were 0.116 MeV for the training set and 0.161 MeV for the test set, as shown in Tab.~\ref{tab4}. The increase is attributed to the unknown theoretical error from newly added samples. Despite these errors, including these samples remains essential to determine whether the model can learn and reproduce the missing experimental magic number.

The predicted neutron ($S_n$) and proton ($S_p$) separation energy were calculated based on the predicted nuclear masses from FCNN which was configured with 60 neurons in the first hidden layer and 30 neurons in the second hidden layer. The results reveal significant oscillations in $S_n$ and $S_p$, reflecting the parity of neutron and proton number, indicative of pairing effects in atomic nuclei, as shown in Fig.~\ref{fig7} (a) and (b). To highlight the impact on the magic number more clearly, we also calculate their rates of change, presented in Fig.~\ref{fig7} (c) and (d). Abrupt changes near the magic number indicate the enhanced stability of magic nuclei. Figure~\ref{fig7} overall shows that the predicted $S_n$ and $S_p$, along with their rate changes $\Delta S_n$ and $\Delta S_p$, exhibit pronounced variations align with the classical neutron magic number (28, 50, 82, 126) and proton magic number (28, 50, 82). Furthermore, similar behaviors are observed near the newly proposed neutron magic number (184, 198) and proton magic numbers (114, 120, 126). We also evaluated the predicted $S_n$, $S_p$, $\Delta S_n$, and $\Delta S_p$ using Method I and Method II with the same data samples. All three methods lead to consistent conclusions, and in regions where experimental data is available, the results match well with the experimental observations.

In Case A, we trained and optimized two models that exhibited strong performance within well-studied regions of the nuclear chart. However, while machine learning methods often outperform traditional theoretical approaches within the training datasets, their predictive accuracy tends to diminish significantly in extrapolated regions. To assess the two methods incorporating physics inputs, we evaluated their extrapolation abilities beyond the training set in a statistically robust and physically meaningful manner, as well as their capability to capture deeper physics principles. We selected four isotopic chains, spanning regions with experimental data to those without. Using two methods, we predicted the binding energies of these isotopes and compared the results with other theoretical approaches. The predictions for four isotopic chains with $Z=$ 28, 50, 82, and 92, as obtained by Method I, Method II, Method III, WS4, RCHB, and FRDM2012, are shown in Fig.~\ref{fig8}.

For the $Z=$ 28 isotopic chain, when extrapolated to $N=$ 55, the predictions from all methods show negligible differences. However, for $N$ $>$ 55, the predictions of Method I, II, III, and FRDM2012 align closely. 
For the $Z=$ 50 isotopic chain, when extrapolated to $N=$ 95, the methods again show minimal differences. When 90 $< N <$ 95, Method III shows slight deviations from the predictions of the other models. For 95 $<N<$ 100, the predictions from Method II and WS4 remain closely aligned. Beyond N=100, the predictions from Method III and WS4 remain closely aligned, whereas Method I begins to diverge significantly from the other models. Although the deviation is less pronounced than that of RCHB.
For the $Z$ = $82$ isotopic chain, when extrapolated to $N=$ 145, the results from all methods are similarly consistent. Beyond $N=$ 145, Method I's predictions diverge noticeably from those of all other models, while Method II and III maintain the same trend as the other theoretical models. Notably, Method III's predictions exhibit excellent agreement with those of WS4. For the $Z=$ 92 isotopic chain, when extrapolated to $N=$ 160, all models except RCHB produce predictions with minor differences. However, for $N$ $>$ 160, Method I's predictions deviate from all other methods, while Method III's predictions align closely with those of RCHB and maintain a smaller gap compared to the other approaches, especially relative to Method I. For the above four isotopic chains, the experimental data used for training are only provided up to $N=$ [45, 85, 133, 148] respectively. while lacking experimental data, the three machine learning methods are still able to align with certain theoretical predictions, demonstrating the extrapolation capability of the physics-informed neural networks.

These results highlight that three distinct Methods demonstrate extrapolation capabilities that are comparable to theoretical models within a certain range. However, the extrapolation capability of Method II and III is notably superior to that of Method I in larger neutron regions. In $Z=$ 92, Method III performs better than the other two models, while it's close to the results predicted by Method II in other $Z$ regions. Over the range of neutron numbers considered in this study, the performance of Method II matches that of the mainstream theoretical models. To the extent that, if the curve labels were removed, it would be challenging to distinguish the machine learning predictions from those of the theoretical models. Our research further demonstrates that the physics-informed and rigorously validated FCNN model, which has the potential to pass a Turing Test, exhibits performance in nuclear mass predictions that is comparable to physicists, highlighting its utility as a robust tool in this field.

\section{Summary and outlook}
\label{sec:summary}
In summary, this study highlights the potential of FCNNs in atomic mass predictions by combining machine learning techniques with domain-specific physics insights. Three distinct methods were developed: Method I focused on leveraging physics principles as input features to directly represent binding energies. Method II employed a macroscopic-microscopic framework with multi-output training to address complex components like deformation and microscopic corrections. Method III is an extension of Method II with additional refinements, including deformation effects modeled through parameter adjustments, to further enhance the predictive accuracy. Methods II and III demonstrated good predictive accuracy, and extrapolation capabilities, as well as the ability to reproduce key physics phenomena such as nucleon pairing effect and magic number. Sensitivity and ablation analyses confirmed the physics interpretability of specific input features, even those with minimal statistical impact. By incorporating predictions from the WS4 model, three methods validated new magic number and provided insights into superheavy nuclei. This work also further compared the mass predictions of all the methods for isotopic chains in regions with sparse experimental data against other theoretical approaches. This study offers a unified framework to advance understanding of nuclear masses and their role in the nuclear landscape.

In the future, machine learning could pave the way for more sophisticated hybrid approaches integrating with nuclear models and experimental analysis. We can extend the studies to encompass diverse nuclear phenomena, improve extrapolation accuracy with innovative network architectures, and explore its broader applications combining with Baysesian emulators~\cite{Xie2023,Zhou:2024hib,Alhassan:2024wrt,Alqahtani:2024ejg}, including hypernuclei properties~\cite{STAR:2019wjm}, nuclear structure and initial conditions of quark-gluon plasma (QGP) in relativistic heavy-ion collisions~\cite{STAR:2024wgy,Jia:2022ozr,Zhang:2021kxj,Giacalone:2021udy,Pang:2019aqb,Yang:2022rlw,Giacalone2024,Huang:2018fzn,Pang:2016vdc,Li:2022ozl}. By bridging machine learning and traditional physics-driven approaches, we expect such explorations would propel the physics capabilities of understanding the properties of QGP in theoretical frameworks and experimental techniques at the Relativistic Heavy-Ion Collider (RHIC) and the Large Hadron Collider (LHC)~\cite{He:2023zin,Zhou:2023pti}.

\begin{acknowledgements}
We would like to thank Wanbing He and Simin Wang for the insightful discussions. This work is supported in part by the National Key Research and Development Program of China under Contract No. 2022YFA1604900, 2024YFA1612600, the National Natural Science Foundation of China (NSFC) under Contract Nos. 12205051, 12025501, 12147101, the Natural Science Foundation of Shanghai under Contract No. 23JC1400200, the STCSM under Grant No. 23590780100, Shanghai Pujiang Talents Program under Contract No. 24PJA009, China Postdoctoral Science Foundation under Grant No. 2024M750489, DOE Research Grant Number DE-SC0024602.
\end{acknowledgements}
\bibliography{ref}
\end{document}